\documentclass[fleqn,usenatbib]{mnras}

\usepackage{newtxtext,newtxmath}
\usepackage{xcolor}
\usepackage{float}

\usepackage[T1]{fontenc}
\usepackage{ae,aecompl}

\usepackage{graphicx}	
\usepackage{amsmath}	
\usepackage{amssymb}	

\title[Cut-off periods in a solar stratified atmospheres]{Slow magnetoacoustic gravity waves in an equilibrium stratified solar atmosphere: cut-off periods through the transition region}

\author[Costa et al.]{
A. Costa$^{1,2}$\thanks{E-mail: acosta@oac.unc.edu.ar}, M. Schneiter$^{1,2,3}$, E. Zurbriggen$^{1,2,4}$
\\
$^{1}$Consejo Nacional de Investigaciones Cient\'\i ficas y T\'ecnicas (CONICET), Argentina.\\
$^{2}$Instituto de Astronom\'\i a Te\'orica y Experimental (IATE), C\'ordoba, Argentina.\\
$^{3}$Universidad Nacional de C\'ordoba (UNC), Facultad de Ciencias Exactas, F\'\i sicas y Naturales, C\'ordoba, Argentina.\\
$^{4}$Universidad Nacional de C\'ordoba, Observatorio Astron\'omico de C\'ordoba (OAC), C\'ordoba, Argentina.
}

\date{Accepted XXX. Received YYY; in original form ZZZ}

\pubyear{2018}

\begin{document}
\label{firstpage}
\pagerange{\pageref{firstpage}--\pageref{lastpage}}
\maketitle

\begin{abstract}

Assuming the thin flux tube approximation, we introduce an analytical model that contemplates the presence of: a non-isothermal temperature; a varying magnetic field and a non-uniform stratified medium
in hydrostatic equilibrium due to   a constant gravity acceleration.
 This allows the study of slow magnetoacoustic 
cut-off periods across the solar transition region, from the base of the solar chromosphere  to the lower corona. The
used temperature profile approaches the VAC solar atmospheric model. The periods obtained are consistent with  observations. 
Similar to the acoustic cut-off periods, the resulting magnetoacoustic gravity ones follow the sharp temperature profile, but  shifted towards larger heights; in other words, at a given height the magnetoacoustic cut-off period is significantly lower than the corresponding acoustic one.
Along a given longitude of an inclined thin magnetic tube, the greater its inclination the softener the temperature gradient it crosses.   
Changes in the   magnetic field intensity do not  significantly modify the periods at the coronal level but  modulate the values below the transition region within periods between $\sim [2\,- 6]\,$min. Within the limitations of our model, we show that monochromatic oscillations of the solar atmosphere are the atmospheric response at its natural frequency to random or impulsive perturbations, and not a consequence of the forcing from the photosphere.

\end{abstract}




\section{Introduction}
The presence of slow MHD  modes  is well established in all parts of the solar atmosphere based on a large amount of observational evidences obtained during the last few decades. They are known to be possible sources for the coronal heating and the solar wind acceleration and to provide important plasma parameter data at the different atmospheric levels through seismological analysis and techniques (see, e.g.,  the review \citet{2016SSRv..200...75N}).      

The dominant photospheric oscillation frequencies observed are known to be global acoustic-gravity waves with periods of approximately five minutes, e.g. \citep{1970ApJ...162..993U}; \citep{gizon2005}.  At  chromospheric levels  the three minute oscillations are dominant (e.g. \citet{2012A&A...539A..23S}), however both, three and five minutes periods, are also observed at coronal layers \citep{2012RSPTA.370.3193D}.   Moreover, observational wave studies of sunspots revealed the presence  of a wealth of alternative periods, from a few seconds to an hour \citep{jess2013}.

The different outward levels present obstacles to the wave propagation such as the rising temperature gradients at the chromospheric and transition layer, the atmospheric density stratification, the location of the equipartition layer  where the gas pressure is equal to the magnetic pressure $\beta=1$.  Due to the changes of the atmospheric properties, waves  are
also modified suffering intensity changes, mode conversions, refractions and reflections. For instance,  chromospheric three
minute oscillations were associated to cavity modes trapped by
the steep temperature gradient at the transition layer and were
also  explained as the response of the solar chromosphere at its
natural frequency to propagating acoustic waves  
\citep{1991A&A...250..235F}. A direct modeling of the chromospheric natural frequencies for different solar atmospheric cases is presented in  \citet{botha2011}. Also,  five minute frequency
observations were explained as the reduction of the acoustic cut-off frequency due to  wave  channeling along inclined magnetic
field lines into the chromosphere  in facular regions  
\citep{2006ESASP.624E..16J}. In the corona, the appearance
of five and three minute propagating compressive waves   were also explained as the
channeling of photospheric and chromospheric perturbations along
the inclined field lines of  magnetic structures.  

Since \citet{1932hydr.book.....L} showed that the acoustic cut-off frequency can be interpreted as  the natural response of the atmosphere to any  disturbance such that waves with lower frequencies become evanescent and cannot transfer energy upwards in a stratified media, extensive literature investigating the propagation and confinement of typical solar modes become possible. In particular, it was established that the chromospheric three minute oscillation can be explained as the natural response (the normal mode given by the cut-off frequency) of the chromosphere to acoustic perturbations \citep[see][]{fleck1991}. 

More recently \citet{afanasyev2015} analyzed the slow magnetoacoustic cut-off modification  due to the presence of magnetic fields in an isothermal stratified solar corona  imposed by the guided field-aligned  plasma dynamic and the gravity action.  They showed that  resulting long wavelength slow magneto-acoustic perturbations -known also as tube modes, due to the field aligned confinement- at the local cut-off frequency could explain the abundant  coronal observations of long period compressive waves ($15-60\,)$min.  

In this work, following the \citet{afanasyev2015} procedure we are interested in adding to the magnetic field consideration  the dispersion effects of  chromospheric regions associated with the usual  temperature gradients that will result in  the cut-off frequency effect.

\section{Wave equation}

To be able of analyzing the  cut-off period in a stratified atmosphere with a varying temperature the ideal thin magnetic flux tube set of linear magnetohydrodynamic equations are employed together with some assumptions that allow an analytic solution. An important consideration is the fact that most of the atmospheric magnetic flux is confined in the form of discrete field tubes with circular cross-section \citep{solanki2006}.

So the axisymmetric flux tube is assumed to be  untwisted, and non-rotating, lacking a steady plasma flow. \citet{afanasyev2015} assumed that the tube was filled with plasma in constant temperature. This assumption is possible because they restricted their attention to the corona only. In this work we will relax this assumption and allow the temperature to vary, allowing us to study the cut-off period in the transition region. Still, the plasma inside the tube is stratified as response to the existing gravitational field that acts in the vertical direction.  The thin flux tube approximation implies that the slow magneto-acoustic wavelength considered is much longer than the tube radius.  This assumption  together with a divergence free evolution is ensured considering a linear variation of the radial magnetic component along the radial direction, a constant axial component of the magnetic field and a parallel wavelength of the perturbations  much longer than the tube radios \citep{zhugzhda1996}.

The set of equations that describe an atmosphere in hydrodynamic equilibrium with the aforementioned assumptions is:

\begin{equation}
\label{eq:1}
  \begin{aligned}
  \rho \left( \frac{\partial u}{\partial t} + u \frac{\partial u}{\partial z} \right) &= - \frac{\partial p}{\partial z} - \rho g, \\
  \frac{\partial s}{\partial t} + u \frac{\partial s}{\partial z} &= 0, \\
  p + \frac{B^2}{8 \pi} &= p^{e}_T, \\
  \frac{\partial B}{\partial t}+ u \frac{\partial B}{\partial z} + 2 B v &=0,   \\
     \frac{\partial \rho}{\partial t} + 2 \rho v + \frac{\partial}{\partial z} (\rho u) &=  0 \\
     p &= p(\rho, s)
  \end{aligned}
\end{equation}
where $\rho$ is the plasma density, $u$ is the longitudinal component of the plasma velocity, $v$ is the radial derivative of the radial component of the plasma velocity, $p$ is the plasma pressure, $s$ the plasma specific entropy, $B$ is the longitudinal component of the magnetic field,  $p^{e}_T$ is the external pressure (assumed constant), and $g$ is the solar gravity acceleration. 

These equations (Eq. \ref{eq:1}) are linearly perturbed around the equilibrium state as:

\begin{equation}
\label{eq:2}
  \begin{aligned}
  \rho &=& \rho_0 + \rho_1,  \; \;\; p &= p_0 + p_1, \; \;\;  &s &= s_0 + s_1  \\
     B &=& B_0 + B_1,        \; \;\; v &= v_1,       \; \;\;  &u &= u_1
  \end{aligned}
\end{equation}

\noindent where subscript $0$ stands for equilibrium state, and $1$ for the departure from it. 

Substituting Eq.(\ref{eq:2}) into Eq.(\ref{eq:1}), keeping the linear terms, after some math work we can reduce the equations into the following form:

\begin{equation}
\label{eq:3}
\frac{B_0}{4 \pi} \frac{\partial B_1}{\partial t} +c_0^2 \frac{\partial \rho _1}{\partial t}-p_0 (\gamma -1 ) R_g u \frac{d s_0}{d z}=0 
\end{equation}

\begin{equation}
\label{eq:4}
\rho_0 \frac{\partial u}{\partial t} + \rho_1 g - \frac{B_0}{4 \pi} \frac{\partial B_1}{\partial z} - \frac{B_1}{4 \pi} \frac{d B_0}{d z}=0
\end{equation}  

\begin{equation}
\label{eq:5}
\begin{aligned}
\rho_0 \frac{\partial B_1}{\partial z} +  \rho_0 \frac{d B_0}{d z} u- B_0 \frac{\partial \rho_1}{\partial t} -  B_0 \frac{d \rho_0}{d z} u - B_0 \rho_0 \frac{\partial u}{\partial z}=0  
\end{aligned}
\end{equation}  
where $c_0$ is the sound speed:

\begin{equation}
c_0^2 = \frac{\gamma P}{\rho}=\frac{\gamma R_g T}{\mu}
\end{equation}
$R_g$, $\gamma=5/3$, and $\mu$\footnote{$\mu=0.635$ for a fully ionized plasma composed of approximately 90\% H + 10\% He.} are the gas constant, the adiabatic index and the average atomic weight, respectively. The temperature $T$ varies with height ($z$).

If we consider a non-isothermal solar atmosphere ($T\equiv T(z)$) in hydrostatic equilibrium,  differentiate Eq.(\ref{eq:4}) with respect to $t$, and Eq.(\ref{eq:5}) with respect to $z$, and rearrange them with all terms corresponding to $u$ and its derivatives together, the following wave equation describing the dynamics in thin magnetic flux tubes   is obtained as:

\begin{equation}
\label{eq:6}
\begin{aligned}
\frac{\partial^2 u}{\partial t^2} -
c_T^2 \frac{\partial^2 u}{\partial z^2}                       &+
\\
\frac{\partial u}{\partial z}  \left[  \frac{1}{B_0} \frac{d B_0}{d z} c_T^2  \frac{V_A^2 - c_0^2}{V_A^2 + c_0^2}  +  \gamma g \frac{c_T^4}{c_0^4} + \begingroup\color{blue}{}  \frac{2 c_T^2}{c_0^2 }        \frac{\gamma-1}{\gamma} c'_0 \endgroup \right]               &+
\\
u \left[  c_T^2 \frac{1}{B_0} \frac{d^2 B_0}{d z^2} +  c_T^2  \frac{c_0^2 -V_A^2 }{c_0^2 + V_A^2 }     \frac{1}{B_0^2} \left( \frac{d B_0}{d z} \right)^2 \right.    &+ 
\\
\frac{1}{B_0}  \frac{d B_0}{d z}  g \frac{c_T^4}{c_0^4} \left( 3 \frac{c_0^2}{V_A^2} + 1 - \gamma \right)    +   
  \frac{c_T^4}{V_A^2} \left( \frac{n^2}{c_T^2} + \frac{g}{c_0^2 h}  \right) & + 
\\
 \begingroup\color{blue}{}\frac{1}{B_0}  \frac{d B_0}{d z}  4 \frac{c_T^4}{c_0 V_A^2} \frac{\gamma -1}{\gamma}  c'_0 \endgroup   
 + \begingroup\color{blue}{}
2 g \frac{c_T^4}{c_0} \left( \frac{1}{\gamma c_T^2 V_A^2} 
+ \frac{1-\gamma}{c_0^4} 
\right)  c'_0  \endgroup      & +  
\\
\left.
\begingroup\color{blue}{}
2 \frac{\gamma-1}{\gamma} \frac{c_T^2}{c_0^2}  c'^2_0 + 
2 \frac{\gamma-1}{\gamma} \frac{c_T^2}{c_0^2}  
c''_0 \endgroup
\right]  & = 0   
\end{aligned}
\end{equation}
where $c'_0=d c_0/dz$ and $c''_0=d^2 c_0/dz^2$. 
$V_A=B_0/\sqrt{4 \pi \rho_0}$ is the  Alfv\'en speed of the equilibrium, $c_T=c_0 V_A/\sqrt{c_0^2+V_A^2} $ is the equilibrium tube speed. The  Brunt-V\"ais\"al\"a frequency $N$ and the scale height $H$ for the non-isothermal case are 
\begin{equation}
\label{eq:8}
N^2= g \left(\frac{1}{\gamma p_0} \frac{d p_0}{d z}  -  \frac{d \rho_0}{d z} \right)= \frac{g^2}{c_0^2} \left(\gamma -1 + \frac{(c_0^2)'}{g} \right)=n +2g^2 \frac{c_0'}{c_0} 
\end{equation}
and 

\begin{equation}
\label{eq:9}
\frac{1}{H}=-\frac{1}{\rho_0} \frac{d \rho_0}{d z} - \frac{1}{T} \frac{d T}{d z} =-\frac{1}{\rho_0} \frac{d \rho_0}{d z} - \frac{1}{c_0^2} \frac{d c_0^2}{d z}=  \frac{1}{h}  -  2 \frac{c_0'}{c_0}
\end{equation}
respectively.  Where $n$ and $h$ in Eq.(\ref{eq:6}) are defined as in \citet{afanasyev2015}. Notice that even though $n$ and $h$ have the same functional form as in equation (5) in  \citet{afanasyev2015} they are no longer isothermal as can be seen in Eq.(\ref{eq:8}) and Eq.(\ref{eq:9}). 

Also, to go from Eqs.(\ref{eq:3}-\ref{eq:5}) to Eq.(\ref{eq:6}) the specific entropy expression was
\begin{equation}
\label{entropy}
ds_0 = c_p \frac{dT_0}{T_0} -R_g \frac{dp_0}{p_0}
\end{equation}
where $c_p=R_g/(\gamma -1)$ is the specific heat coefficient at constant pressure. It is clear from the expression Eq.(\ref{entropy}) that for a constant temperature one recovers the entropy employed in \citet{afanasyev2015}.

The colored parts (blue for the on-line version) correspond to the additional terms connected with the temperature non-uniformity. The way we arranged the equation is such that it matches the form obtained for the isothermal solution introduced in the work of \citet{afanasyev2015} plus something else that corresponds to the full solution. Notice that, when considering a constant temperature ($T_0(z)=T_0$) the sound speed becomes constant, and all its derivatives vanish and consequently the colored terms in Eq.(\ref{eq:6}). Thus,  Eq.($4$) of \citet{afanasyev2015} is recovered. If, in addition,  we  assume
a constant $B_0$,  the resulting Eq.(7) exactly matches
Eq.(3.9) of Roberts (2006).

 As stated  by \citet{afanasyev2015},  if $B_0$ is constant, 
Eq.($4$) of their paper completely coincides with Eq.(3.9) of \citet{2006roberts}. However,  to obtain Eq.(3.9) in \citet{2006roberts} the temperature was not explicitly assumed constant (as shown by the formal dependence with height of the sound speed and the pressure scale height) although it reduces to an isothermal case.

\section{The cut-off period}

From Eq.(\ref{eq:6}), and with the aim of writing it in the Klein Gordon form, we identify the coefficients $K_1$ and $K_2$ as:

\begin{equation}
K_1 =  \frac{1}{B_0}  \frac{d B_0}{d z} c_T^2 \frac{V_A^2 - c_0^2}{V_A^2 + c_0^2} + g \gamma \frac{c_T^4}{c_0^4} +\begingroup\color{blue}{} 2 \frac{c_T^2}{c_0} \frac{\gamma - 1}{\gamma} c'_0 \endgroup 
\end{equation}

\begin{equation}
\begin{aligned}
K_2 &=  c_T^2 \frac{1}{B_0} \frac{d^2 B_0}{d z^2} +  c_T^2  \frac{c_0^2 -V_A^2 }{c_0^2 + V_A^2 }     \frac{1}{B_0^2} \left( \frac{d B_0}{d z} \right)^2     
\\
&+\frac{1}{B_0}  \frac{d B_0}{d z}  g \frac{c_T^4}{c_0^4} \left( 3 \frac{c_0^2}{V_A^2} + 1 - \gamma \right)    +   
  \frac{c_T^4}{V_A^2} \left( \frac{n^2}{c_T^2} + \frac{g}{c_0^2 h}  \right)  
\\
& + \begingroup\color{blue}{}\frac{1}{B_0}  \frac{d B_0}{d z}  4 \frac{c_T^4}{c_0 V_A^2} \frac{\gamma -1}{\gamma}  c'_0 \endgroup   
 + \begingroup\color{blue}{}
2 g \frac{c_T^4}{c_0} \left( \frac{1}{\gamma c_T^2 V_A^2} 
+ \frac{1-\gamma}{c_0^4} 
\right)  c'_0  \endgroup        
\\
& +\begingroup\color{blue}{}
2 \frac{\gamma-1}{\gamma} \frac{c_T^2}{c_0^2}  c'^2_0 + 
2 \frac{\gamma-1}{\gamma} \frac{c_T^2}{c_0^2}  
c''_0 \endgroup
\end{aligned}
\end{equation}

Thus, we can reduce the equation to the Klein Gordon form for the variable $U$

\begin{equation} 
\frac{\partial^2 U}{\partial t^2} - c_T^2 \frac{\partial^2 U}{\partial z^2} + \omega_u^2 U =0   
\end{equation}
where $u=\exp^{\psi(z)} U(z)$ and  $\psi'= K_1/(2 c_T^2)$.
Thus the square of the cut-off frequency is 
 
\begin{equation}
\omega_u^2 = -c_T^2 \psi'^2 - c_T^2 \psi'' + K_1 \psi'' + K_2 
\end{equation}

It is straight forward to show that if we assume a constant sound speed we recover the solution obtained in \citet{afanasyev2015}, since all the terms with derivatives
of the sound speed vanish. Thus, the hydrodynamic cut-off frequency is also obtained  when the zero magnetic field limit and constant temperature are considered (see Eq.($11$) in \citet{afanasyev2015}), and the same is true for the case of constant magnetic field (see Eq.($12$) in \citet{afanasyev2015}).  

\section{The temperature model}
\label{temperature}

Now that we have an analytical model that allows the inclusion of variations in temperature, we can propose a profile  to describe the transition region such as:

\begin{equation}
T_0(z) = a_0 \tanh (z-a_1)/a_2+a_3
\end{equation}
where $t_s$ is the variable that emulates the solar atmospheric temperature profile, $a_{0,1,2,3}$ are parameters that allow the adjustment to the desired values and shape.  
With the proper choice of these parameters the profile matches the chromospheric  temperature and the coronal temperature as proposed in \citet{vernazza1981}. 
To study the cut-off variation with temperature we choose two temperatures, 
$T(z=0 \,Mm)=1\times 10^4 \,K$ and $T(z=8\,Mm)=1.4 \times 10^6 \,K$, where $z=0$ represents the base of the chromosphere.

\begin{figure}
	\includegraphics[width=\columnwidth]{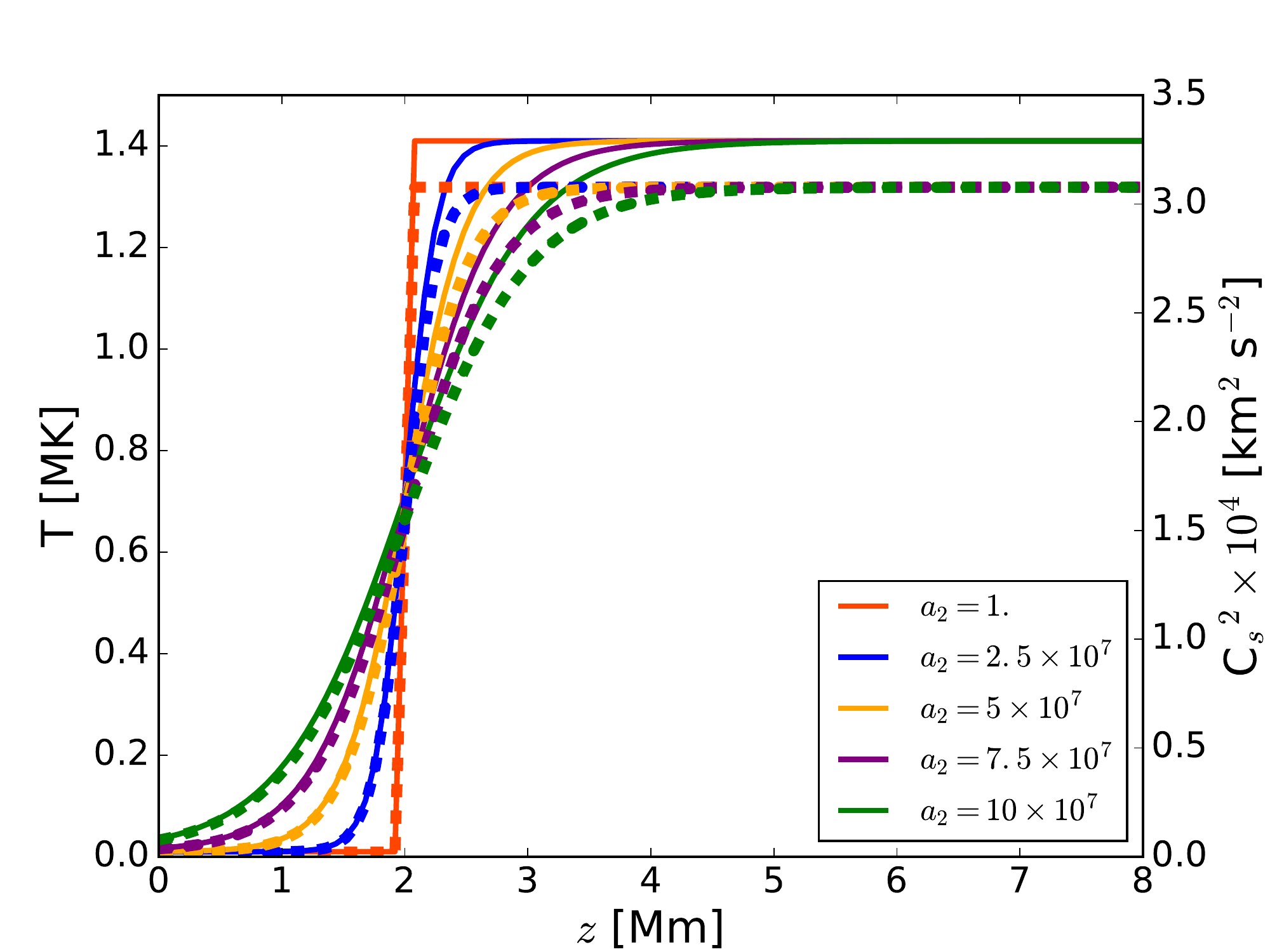}
    \caption{(Full lines) Temperature profile as a function of height $z$ for different $a_2$ values, going from a sharp transition region to a soft one. All of them begin at the base of the chromosphere and extend to the beginning of the corona. (Dashed-lines) Corresponding sound speed as a function of height. $a_0= 7\times 10^5\,$K, $a_1=2 \,$Mm and $a_3=7.1 \times 10^5$K.}
    \label{fig:temperature-profile}
\end{figure}

By fixing the parameters $a_0= 7\times 10^5\,$K, $a_1=2 \,$Mm and $a_3=7.1 \times 10^5$K we guarantee the mentioned temperatures at both ends. Varying $a_2$ from low to high values produces a transition region that goes from a sharp  profile, almost discontinuous, to a soft one. Figure \ref{fig:temperature-profile} (full lines) shows this behavior for $a_2$ in the height range $[0-8]\,Mm $.

\section{Model parameters}
\label{Model parameters}

\begin{figure}
	\includegraphics[width=\columnwidth]{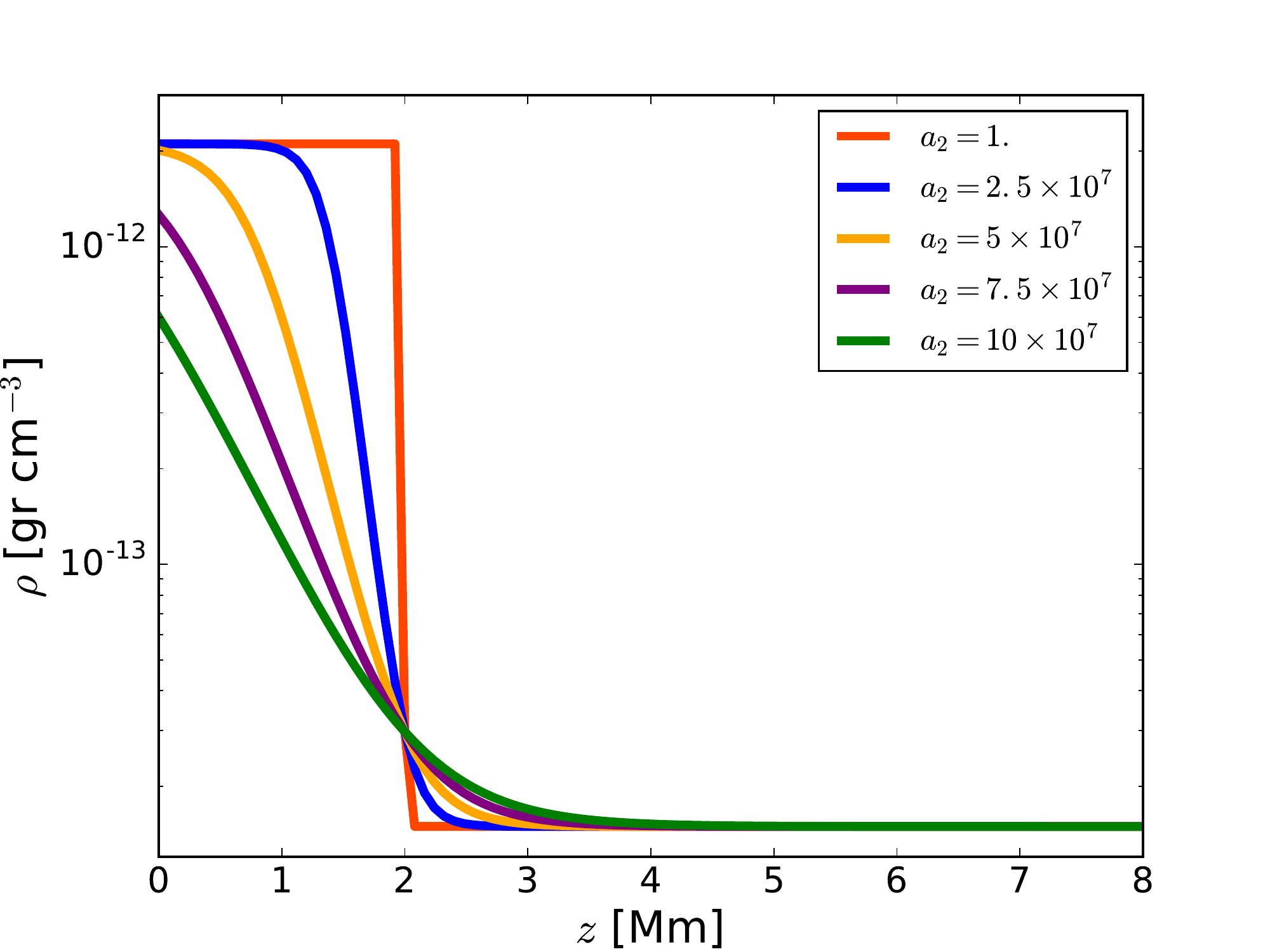}
	\caption{Density profile as a function of height for different $a_2$ values.  $a_0= 7\times 10^5\,$K, $a_1=2 \,$Mm and $a_3=7.1 \times 10^5$K.}
    \label{fig:2}
\end{figure}

The choice of the temperature model determines both the sound speed profile as function of height, since $c_0 = \sqrt[]{\gamma p_0/\rho_0}$ and $p_0=\rho_0 R_g T_0/\mu$,  as shown in  Figure \ref{fig:temperature-profile} (dashed lines) as well as the density profile, shown in  Figure \ref{fig:2}.
The base number density  value is $n_0=1 \times 10^{11}$ cm$^{-3}$.

Following \citet{afanasyev2015} we employed the expression $B(z)=B_0 \exp^{-z/l}$ that corresponds to a divergent magnetic flux tube, where $B_0$ is the magnitude of the magnetic field at the lower boundary and $l$ is the magnetic field scale height. Initially, with the aim of comparing with the hydrodynamic case as well as analyzing  the effect of  the temperature profile, we chose $l=0.2 \,h_{b}$\footnote{$h_b$ is simply the value of $h$ at the base of the chromosphere,  $z=0$.} and $B_0=10\,$G.

\section{Results and discussion}

\begin{figure}
\includegraphics[width=1. \columnwidth]{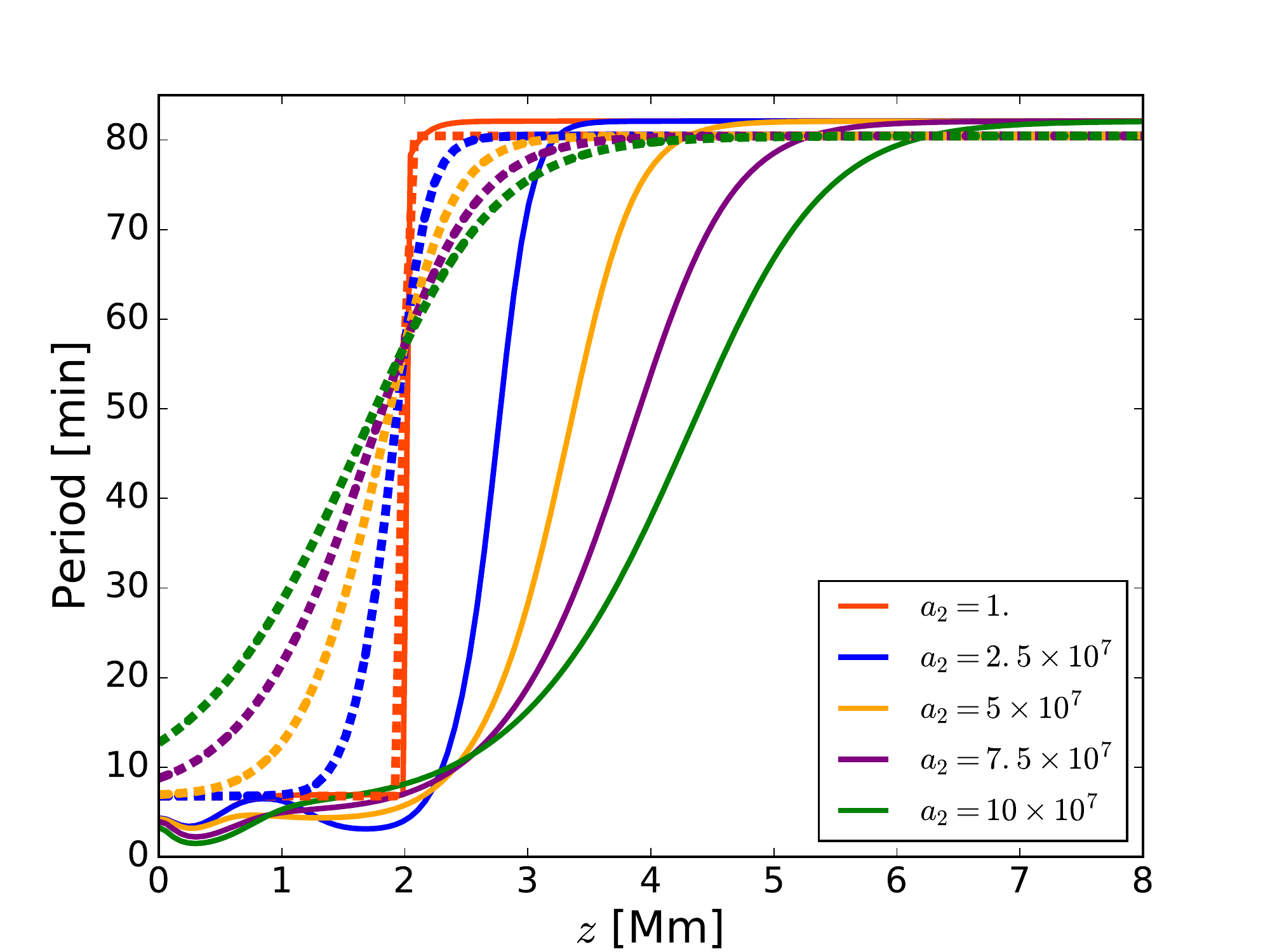}
\caption{(Full lines) Cut-off  periods of magnetoacoustic gravity waves for the different temperature profiles and the magnetic field scale height $l=0.2\, h_b$; (Dashed lines) cut-off period of acoustic waves. $a_0= 7\times 10^5\,$K, $a_1=2 \,$Mm and $a_3=7.1 \times 10^5$K.}
    \label{fig:waves}
\end{figure}

When employing these models to calculate the wave cut-off periods we get the results shown in Figure \ref{fig:waves} where we analyze both, the pure acoustic case ($\omega_a=\gamma g/4 \pi c_0$)\footnote{obtained from Eq.(\ref{eq:6}) describing the dynamics in the thin magnetic flux tube for the  infinite magnetic field limit (see e.g. \citet{afanasyev2015}).} and the magnetoacoustic  one for different shapes of the transition region as explained in Sec.\ref{temperature}. 

From Figure \ref{fig:waves} we can see that the magnetoacoustic periods (full lines in Figure \ref{fig:waves}) go from $\sim 4\,$min at  the base of the chromosphere to $\sim 82\,$min at the corona, consistent with \citet{afanasyev2015}. If we look close to the base of the chromosphere it is possible to see that the periods start at values near $4\,$min, then experience a decrease to $\sim 2-3\,$min at $0.3\,$Mm, and later increase again to values of $\sim 4-6\,$min at heights of $\sim 0.5-1\,$Mm, to later start increasing slow-fast, depending on the temperature model, until reaching the final value of $\sim 82\,$min. Hence, the temperature transition region sharpness produces a similar behavior on the cut-off period but shifted to higher $z$ values with respect to the acoustic cut-off periods. This result is consistent with the observational multi-height magnetoacoustic cut-off frequencies obtained by \citet{yuan2014}.
When comparing the pure acoustic cut-off (dashed lines in Figure \ref{fig:waves}) with the magnetoacoustic gravity wave cut-off we see that at the base of the chromosphere the starting point is larger, from $\sim 5\,$min to $\sim 13\,$min, increasing monotonically until reaching the $80\,$min cut-off value at the corona \citep{afanasyev2015}. 

\begin{figure}
\includegraphics[width=1.1 \columnwidth]{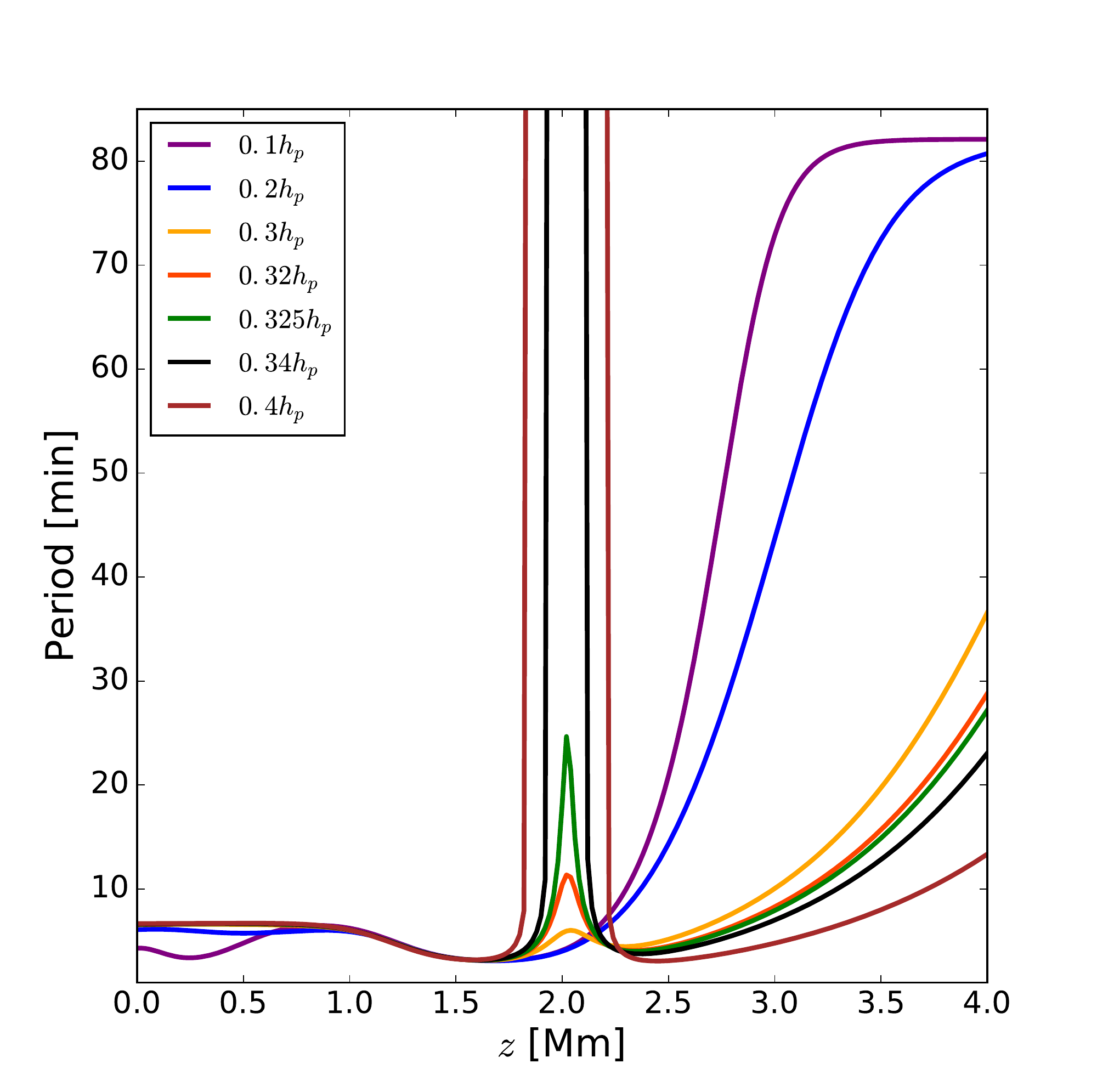}
\caption{Cut-off periods corresponding to the temperature profile with  $a_2=2.5 \times 10^7$ and different inclination given by the parameter  $l=[0.1, \ 0.2, \ 0.3, \ 0.32, \ 0.325, \ 0.34, \ 0.4]h_p$.}
\label{fig:gap}
\end{figure}

Up to this point the models contemplate a starting magnetic field value of $B_0=10\,$G with a given inclination parameter $l$ (see section \ref{Model parameters}). Following, we vary the inclination of the magnetic field with respect to the scale height $h_b$, from $l=0.1 \,h_b$ (large inclination) to $l=100\, h_b$ (almost straight) for the fixed temperature gradient $a_2=2.5 \times 10^7$, which corresponds to a  quiet sharp transition region (see Figure \ref{fig:temperature-profile}). 
In  Figure \ref{fig:gap} we distinguish two differentiated behaviors:  1) when the magnetic field is sufficiently curved (e.g. $l=0,1 \,h_b$) one obtains a smooth behavior whose period increases with temperature as in Figure \ref{fig:waves};  2) larger values ($l=1;\,5;\,10;\,100 \,h_b$) produce a discontinuity or gap centered at the  transition region with gap widths ($w_g$) that go from $\sim 0,8\,$Mm to $\sim 1,9\,$Mm,  corresponding to increasing  $l$ values.
The straighter the magnetic tube,  the greater $w_g$ or window where all cut-off periods are allowed. We attribute this behavior to the influence of the steepening of the temperature gradient along the tube field line. To clarify this point, if the curvature of the magnetic tube is large (small $l$ value) then the temperature gradient transited by the tube is softened.

\begin{figure}
\includegraphics[width=1.1 \columnwidth]{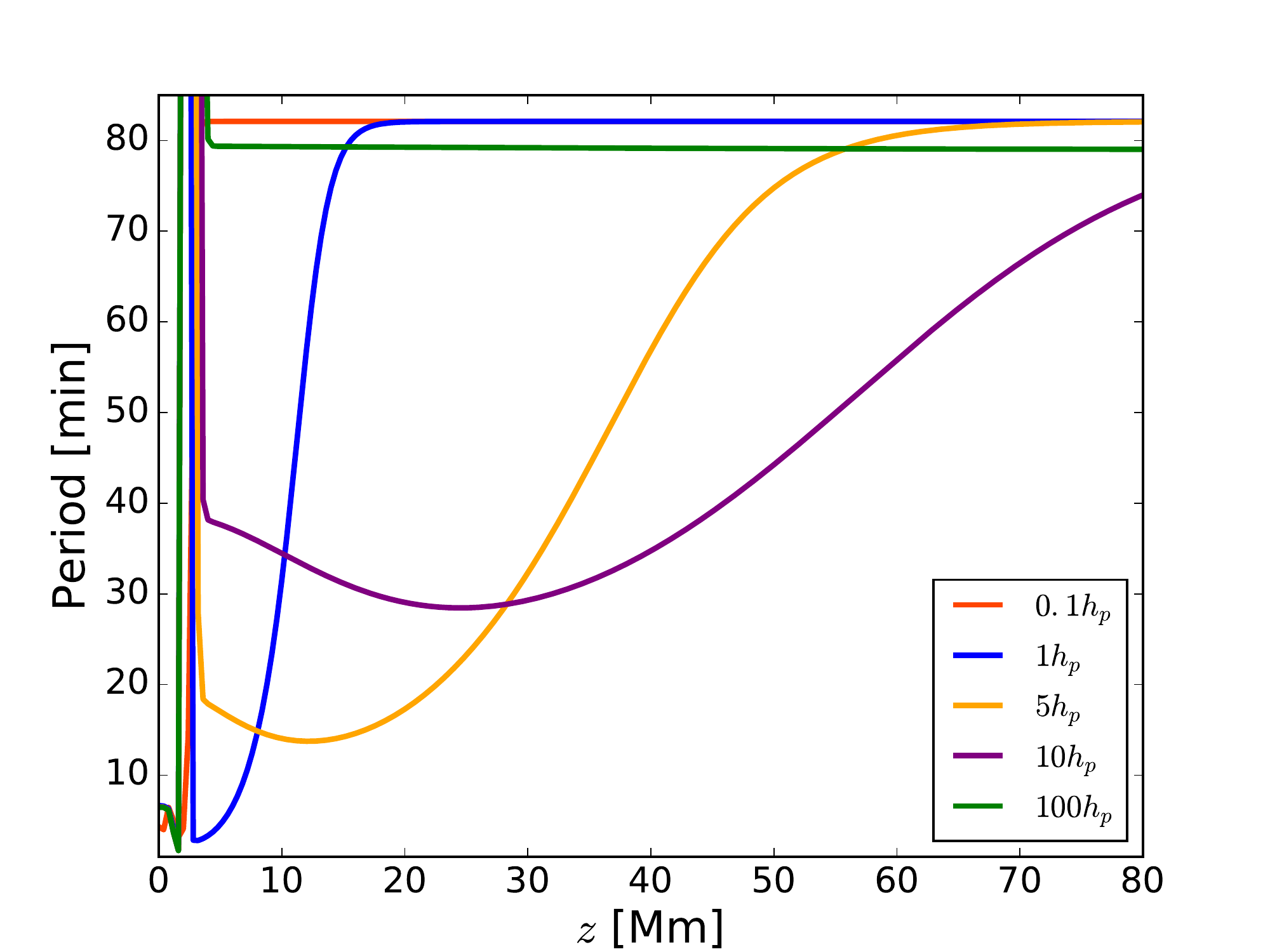}
\caption{Cut-off periods corresponding to the temperature profile with  $a_2=2.5 \times 10^7$, located after the transition region.}
\label{fig:lvar}
\end{figure}

Figure \ref{fig:lvar} is the same as Figure \ref{fig:gap} but reaching larger coronal altitudes. The aim is to analyze the fall of the cut-off period after the gap as a function of the magnetic field inclination as well as to visualize how fast each model converges to its final period. One can notice that, the more inclined magnetic field tube the smaller the drop after the gap. Table \ref{table:I} shows a quantitative description. Notice that, all the curves reach a cut-off period of $\sim 82\,$min ($h_c$ in Table \ref{table:I}), but the straighter the tube the longer the distance it takes to reach the final value.

Looking at height $z=4 \times 10^9\,$Mm in Figure \ref{fig:lvar} we note that the smaller the value of $l$ the larger the cut-off period. This behavior is in agreement with most sunspot observations e.g., \citet{jess2013}, and also with the observational cut-off reconstruction by \citet{yuan2014}, as well as our previous results. That is, the analysis of Figure \ref{fig:waves}, where we see that the magnetoacoustic gravity cut-off periods, in comparison to the pure acoustic periods, are shifted towards larger altitudes. Hence, we can say that the increase of the periods can be attributed to the magnetic tube inclination.

\begin{table}
\begin{center}
\begin{tabular}{ r c c c c}
 $l$       & $w_g$   &$d$              & $h_c$  \\ 
 \hline
 \hline
 0.1 $h_b$ & No gap  &     N/A         &  3.6 Mm \\  
 1   $h_b$ & 0.8 Mm  & $\sim  3$ min  &  20  Mm \\  
 5   $h_b$ & 1.2 Mm  & $\sim 14$ min  &  70  Mm \\  
 10  $h_b$ & 1.5 Mm  & $\sim 28$ min  &  120 Mm \\  
 100 $h_b$ & 1.9 Mm  & $\sim 79$ min  &  600 Mm  
\end{tabular}
\end{center}
\caption{ $l$, $w_g$, $d$ and $h_c$ are the magnetic field inclination parameter, gap width, drop after gap and convergence height, respectively. }\label{table:I}
\end{table}

\begin{figure}
\includegraphics[width=1.1 \columnwidth]{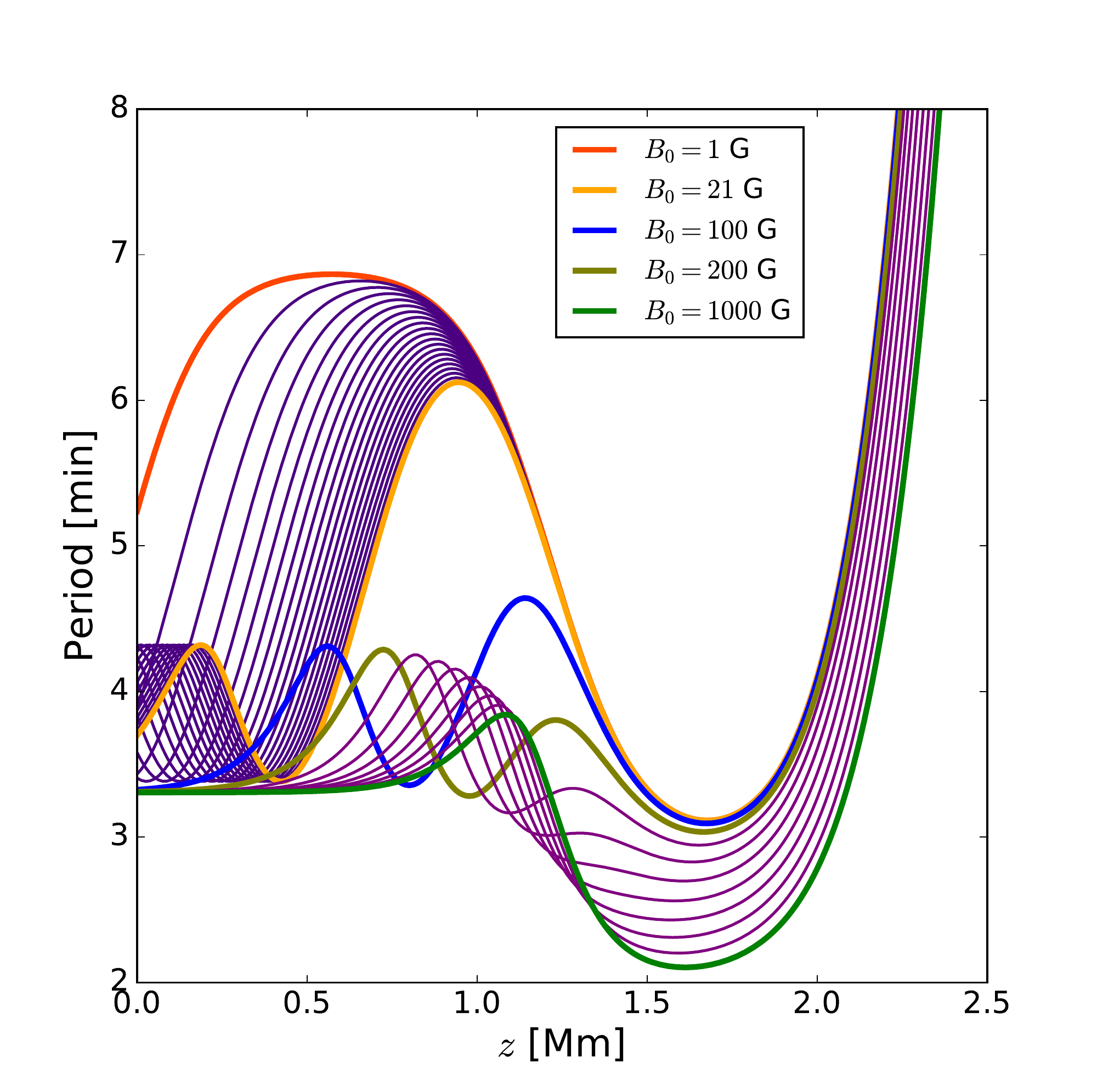}
\caption{Cut-off periods corresponding to the temperature profile with $a_2=2.5\times 10^7$ and several magnetic field intensities, $B_0=[1, \ 2, \ 3, \ 4, \ 5, \ 6, \ 7, \ 8, \ 9, \ 10, \ 11, \ 12, \ 13, \ 14, \ 15, \ 16, \ 17, \ 18, \ 19, \ 20, \ 21,$ $100, \ 200, \ 300, \ 400, \ 500, \ 600, \ 700, \ 800, \ 900, \ 1000]~$G. Purple curves represent all the listed intermediate magnetic intensities not included in the legend.}
\label{fig:bvar}
\end{figure}

In order to analyze the role of the magnetic field intensity ($B_0$) Figure \ref{fig:bvar} shows profiles with $B_0$ that go from $5\,$G to $1000\,$G at the base of the chromosphere for a given curvature ($l=0.1 h_b$). 
A more precise model should consider that the transition region is modified due to the presence of intense magnetic fields. This could be considered qualitatively from the change of the parameter $a_2$. Here, given that the model is linear and analytical, we only estimate the trend behavior for strong  magnetic field values.

Varying the intensity  of the magnetic field we find that  it does not affect the periods in the corona, but has some effects on the  chromosphere, before the temperature transition region. Looking close to this region (see Figure \ref{fig:bvar}), we can observe that small $B_0$ values produce a notable increase in period (up to $\sim 6\,$min). Larger magnetic field values produce a smaller increase, they start at values of $\sim 3\,$min, increase to just above $4\,$min and later fall to $\sim 4\,$min again before starting their climb to the $82\,$min period at the corona. The period bump is shifted towards larger chromospheric heights with increasing magnetic field values. The largest $B_0$ value ($1000\,$G) shows a greater fall, to a period of $\sim 2\,$min, which is consistent with umbra wave observations in sunspots e.g., see \citet{sharma2017}.
Summarizing, we  see that the increase of the magnitude of the magnetic field has no important effect over the periods in the corona, but it can modulate the values before the transition region within periods between $2\,$min to $6\,$min.

\section{Conclusions}
We derived a temperature dependent model for longitudinal waves propagating in a vertically stratified  plasma by the gravity. This model  allowed the study of the cut-off periods in the corona as well as in  the chromosphere and the transition region. We imposed a temperature model that can represent the variation with height of the equilibrium temperature  
to study the way in which  the cut-off periods of slow magnetoacoustic gravity waves are affected by the strong temperature  gradient. We compared the resulting function  of  allowed acoustic periods with the magnetoacoustic ones.  Vertical profiles of both,  acoustic  and slow magnetoacoustic cut-off periods are similar to the equilibrium temperature profile with the transition region, but the magnetoacoustic periods are shifted towards larger heights  resulting in a decrease of the corresponding cut-off period for a given altitude. Also, the extremes of the  period curves approximately  suit the typical chromospheric observed frequencies (see e.g., the review by \citet{2016SSRv..200...75N} and references therein). At coronal altitudes, where the temperature can be assumed constant, our results reproduce the frequencies obtained by \citet{afanasyev2015}.

We then analyzed the effects  of both, the magnetic field inclination and the magnetic field intensity. 
Most observations of  slow modes in active regions have interpreted the change of the oscillatory period -from a few minutes at the umbra to increasing values at the penumbra- as the effect of the inclination  of the  magnetic field that upwardly guide magnetoacoustic gravity waves from the chromosphere to the corona (see e.g., \citet{jess2013}, \citet{yuan2014}, \citet{reznikova2012}).  
If a sunspot can be thought as a bunch of threads of thin magnetic tubes \citep{parker1979}, our model gives account   that the more inclined the magnetic field tube the larger the cut-off period obtained. 

We also varied the magnetic field intensity to study its effect on the  periods.  We find that, while the magnitude of the magnetic field is not determinant for the coronal periods it influences the atmospheric regions below the transition region, i.e., the larger the magnetic intensity the greater the fall of the periods. Consistently with umbral wave observations in \citet{sharma2017} we found that a magnetic field intensity of $B_0=1000\,$G can be associated with  periods of $\sim 2\,$min. 

As discussed in the literature, there are many observational evidences that suggest that the oscillations detected along the different atmospheric layers are produced by  slow magnetoacoustic gravity waves propagating upwards which are altered by the local physical conditions, e.g., temperature variations, gravity stratification, inclination of the magnetic field, and/or magnetic field intensity. However, it is not clear yet whether these characteristic perturbations are  externally driven (e.g., p modes) or locally excited (e.g., random  convective motions). 
If our model is sufficiently accurate, given  its limitations and approximations, we here showed that  the cut-off frequency can be thought of as the natural response of the different atmospheric regions to random perturbations rather than excited by external actions.




\bibliographystyle{mnras}
\bibliography{ref} 








\bsp	
\label{lastpage}
\end{document}